\documentclass[]{emulateapj}

\begin{document}

\newcommand{\kms}{km~s$^{-1}$}
\newcommand{\msun}{$M_{\odot}$}
\newcommand{\rsun}{$R_{\odot}$}
\newcommand{\teff}{$T_{\rm eff}$}
\newcommand{\logg}{$\log{g}$}

\slugcomment{ApJ accepted}

\title{Most Double Degenerate Low Mass White Dwarf Binaries Merge}

\author{Warren R.\ Brown$^1$,
	Mukremin Kilic$^2$,
	Scott J.\ Kenyon$^1$,
	and A.\ Gianninas$^2$
	}

\affil{ $^1$Smithsonian Astrophysical Observatory, 60 Garden St, Cambridge, MA 02138 USA\\
	$^2$Homer L. Dodge Department of Physics and Astronomy, University of Oklahoma, 440 W. Brooks St., Norman, OK, 73019 USA
	}

\email{wbrown@cfa.harvard.edu, kilic@ou.edu}

\shorttitle{ Most White Dwarf Binaries Merge }
\shortauthors{Brown et al.}

\begin{abstract}

	We estimate the merger rate of double degenerate binaries containing
extremely low mass (ELM) $<0.3$ \msun\ white dwarfs in the Galaxy.  Such white
dwarfs are detectable for timescales of 0.1 Gyr -- 1 Gyr in the ELM Survey; the
binaries they reside in have gravitational wave merger times of 0.001 Gyr -- 100
Gyr.  To explain the observed distribution requires that most ELM white dwarf binary
progenitors detach from the common envelope phase with $<$1 hr orbital periods.  We
calculate the local space density of ELM white dwarf binaries and estimate a merger
rate of $3\times10^{-3}$ yr$^{-1}$ over the entire disk of the Milky Way; the merger
rate in the halo is 10 times smaller.  The ELM white dwarf binary merger rate 
exceeds by a factor of 40 the formation rate of stable mass transfer AM~CVn
binaries, marginally exceeds the rate of underluminous supernovae, and is
identical to the formation rate of R~CrB stars. On this basis, we conclude that ELM
white dwarf binaries can be the progenitors of all observed AM~CVn and 
possibly underluminous supernovae, however the majority of He+CO white dwarf
binaries go through unstable mass transfer and merge, e.g.\ into single massive
$\sim$1 \msun\ white dwarfs.

\end{abstract}


\section{INTRODUCTION}

	Double degenerate white dwarf (WD) binaries with orbital periods less than
about 6 hours will merge within a Hubble time due to energy and angular momentum
loss from gravitational wave radiation.  Pairs of normal, 0.6 \msun\ CO WDs are
rarely found in such compact binaries, however.  The Supernovae Progenitor Survey
obtained precision radial velocities for 1,014 nearby WDs and identified 5 double
degenerate binaries that will merge within a Hubble time \citep{napiwotzki07}.  
This result is consistent with simulations:  binary population synthesis models 
predict that most of the WD+WD mergers in the Milky Way are low mass He+He WD and
He+CO WD binaries \citep{iben90, han98, nelemans01b}.  The reason is that
helium-core $\lesssim0.3$ \msun\ ELM WDs form out of common envelope evolution in
ultra-compact binaries \citep{marsh95}, exactly the type of binaries that quickly
merge.

	Over the past five years, our targeted survey for ELM WDs has uncovered 76 
short period $P \le 1$ day double degenerate binaries containing an ELM WD 
\citep{brown10c, brown12a, brown13a, brown16a, kilic10, kilic11a, kilic12a, 
gianninas15}.  The success of the ELM Survey in finding these binaries is explained 
in part by the long evolutionary times of the ELM WDs.  ELM WDs are predicted to 
have thick H envelopes with residual shell burning that cause them to remain hot 
$T_{\rm eff} > 8$,000 K and luminous $M_g<9$ mag for many Gyr \citep{sarna00, 
panei07, althaus13, istrate14}.  The gravitational wave merger times of the observed 
binaries, on the other hand, are as short at 1 Myr \citep{brown11b}.

	Our ELM Survey is now large enough and complete enough that we can use the
distribution of objects to infer the merger rate of ELM WD binaries throughout the
Milky Way.  We made an early attempt at this calculation using the first 12 ELM WDs
found in the original Hypervelocity Star Survey \citep{brown11a}.  This first sample
was based on WDs identified by visual inspection; stellar atmosphere fits
subsequently uncovered additional ELM WDs in the Hypervelocity Star Survey
\citep{brown13a}.  More importantly, our subsequent discoveries include four systems
with orbital periods $<40$ min and with merger times $<27$ Myr \citep{brown11b,
kilic11c, kilic11b, kilic14}.  The existence of such short merger-time systems
implies that ELM WD binary mergers are more frequent than previously thought.

	An outstanding question is what happens when ELM WD binaries merge.  There 
are at least three possible outcomes:  1) a long-lived stable mass-transfer binary, 
2) an explosion, and 3) a merger into a single massive WD \citep[e.g.,][]{webbink84, 
iben90, kilic10}.  The three outcomes depend on the stability of mass transfer and 
thus on the mass ratio of the donor and accretor, however there are many 
complications \citep[e.g.,][]{marsh04, nelson04, kaplan12}. For reference, the 
average ELM WD binary in our sample has a mass ratio of about $M_{\rm 
donor}$:$M_{\rm accretor}=1$:4 and total mass of about 1 \msun\ \citep{andrews14, 
brown16a}.

	\citet{kremer15} argue that essentially all the observed ELM WD binaries
will evolve into stable mass-transfer configurations, binaries in which helium mass
transfer can proceed stably for billions of years.  Observationally, long-lived
stable helium mass-transfer binaries will appear as AM Canum Venaticorum (AM~CVn)
systems \citep{warner95, solheim10}.

	Theoretical models suggest that helium mass transfer may lead to a series of 
thermonuclear flashes, the last of which may result in an underluminous explosion 
dubbed a supernova ``.Ia'' \citep{bildsten07, shen09, waldman11}.  However, models 
of ELM WD binaries that are probable AM~CVn progenitors find that the conditions for 
detonation may never develop \citep{piersanti15}.  The implication is that most ELM 
WD binaries will end up as long-lived AM~CVn systems.

	Alternatively, the result may be a merger.  If the initial binary mass ratio 
is $M_{\rm donor} / M_{\rm accretor} > 2/3$, or if the initial hydrogen mass 
transfer rate is super-Eddington, a common envelope may form and cause the two WDs 
to quickly merge into a single massive WD \citep{webbink84, han99, dan11, shen15}.  
Observationally, the merger event should form a single object and ignite nuclear 
burning, appearing as an extreme helium star, including an R~CrB star 
\citep{paczynski71, webbink84, saio02}, or He-rich sdO star \citep{heber08}.
	We cannot wait long enough to observe the outcome of an ELM WD binary
merger, however we can constrain the outcome by considering the formation rates of
AM~CVn, underluminous supernovae, and R~CrB.

	We begin this paper by presenting a clean sample of ELM WD binaries.  We
divide the sample into disk (63\%) and halo (37\%) objects on the basis of
kinematics and spatial position, and then derive the local space density of disk and
halo ELM WD binaries using standard Galactic stellar density models.  We estimate
merger rates in two ways:  by correcting the observed sample for systems that have
cooled or merged, and by forward modeling to match the observed distributions.  We
conclude that the vast majority of ELM WD binaries must form at short $<$1 hr
orbital periods to explain the observed distribution of ELM WD binaries.  We compare
the estimated ELM WD binary merger rate with the formation rates of AM~CVn systems,
underluminous supernovae, and R~CrB stars, and conclude that the rates favor 
non-explosive merger events.

\section{MODEL INPUT}

\subsection{Clean Sample}

	ELM Survey targets were selected by de-reddened $g$-band magnitudes 
($15<g_0<20$ mag) and by color.  \citet{brown12b, brown12a} publish the exact color 
selection, which was designed to target A- and B-type stars with surface gravities 
between normal main sequence stars with $\log{g} \simeq 4$ and normal hydrogen 
atmosphere WDs with $\log{g} \simeq 8$.  Our observing strategy was to acquire a 
spectrum of each color-selected target, identify its nature, and then re-observe 
each $5 \lesssim \log{g} \lesssim 7$ ELM WD candidate until we constrain its 
orbital solution.


	We define a clean set of ELM WD binaries from the ELM Survey by first
restricting the sample to those binaries with $k>75$ \kms.  Sensitivity tests
demonstrate that our multi-epoch spectroscopy should detect 95\% of binary systems
with semi-amplitude $k=75$ \kms\ and 99\% of binary systems with $k=100$ \kms\ out
to $P=2$ day orbital periods \citep{brown16a}.

	Next, we restrict the sample to those objects with $4.85 < \log{g} < 7.15$, 
a range over which ELM Survey follow-up observations are 95\% complete.  The ELM 
Survey color selection in $(g-r)$ provides a built-in temperature selection of 
8$,$000~K$ < T_{\rm eff} < 22$,000~K.

	Finally, we remove the two ELM WDs that were not drawn from the Sloan
Digital Sky Survey (SDSS) photometric catalog:  NLTT~11748 \citep{kilic10b} and the
LAMOST object J0308+5140 \citep{gianninas15}.  Taken together, these cuts leave us
with a clean sample of 60 ELM WD binaries in a well-defined footprint of sky.

	We present the clean sample in Table \ref{tab:clean} sorted by orbital 
period.  \citet{brown16a} estimate that this ELM WD sample is 60\% complete, given 
the number of ELM WD candidates that have unknown or poorly constrained orbital 
parameters.  We will thus correct all relevant quantities in this paper assuming a 
60\% completeness.

\begin{figure}          
 \includegraphics[width=3.5in]{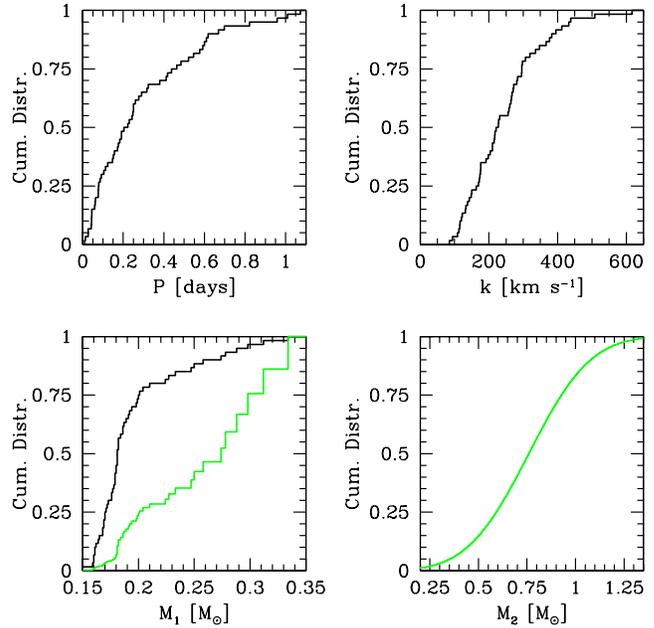}
 \caption{ \label{fig:orbparam}
	Orbital parameter distributions for the clean sample of ELM WD binaries.  
Orbital period $P$ and semi-amplitude $k$ are directly measured.  ELM WD mass $M_1$ 
is derived from spectroscopic \teff\ and \logg\ measurements using \citet{althaus13} 
models; the green line shows the $M_1$ distribution scaled by the inverse of 
evolutionary time.  Secondary mass $M_2$ is inferred from the observations 
\citep{brown16a}. } \end{figure}

\subsection{White Dwarf Parameters}

	We adopt the ELM WD masses and luminosities reported in \citet{brown16a}.  
These parameters were derived by comparing spectroscopic \teff\ and $\log{g}$ 
measurements to the ELM WD evolutionary models of \citet{althaus13}.  A notable 
feature of the models is that $\lesssim$0.18 \msun\ ELM WDs evolve relatively slowly 
due to residual burning in the hydrogen envelopes, while $\gtrsim$0.18 \msun\ ELM 
WDs evolve more quickly due to unstable hydrogen shell flashes.  The ELM WD models 
follow standard WD mass-radius relations except during these moments of unstable 
hydrogen shell burning, when the effective radius of the ELM WD is inflated.

	Quantitatively, the median observed ELM WD in our clean sample has mass 
$M_1=0.18$ \msun, absolute magnitude $M_g=+8.5$, apparent magnitude $g_0=18.5$, and 
heliocentric distance $r=1.0$ kpc.  The SDSS covers a high Galactic latitude region 
of sky; the median object has latitude $b=45\arcdeg$ and vertical distance above the 
Galactic plane $Z=0.7$ kpc.

	We include the ELM WD masses $M_1$ in Table \ref{tab:clean}.  The $M_1$ 
uncertainties are derived by propagating \teff\ and $\log{g}$ errors through the 
tracks and then adding 0.01 \msun\ in quadrature, motivated by the comparison with 
\citet{istrate14} models.  The \citet{istrate14} models yield mass and luminosity 
estimates that agree to within $\pm$6\% of the \citet{althaus13} models, however 
they have much longer evolutionary times due to the absence of gravitational 
settling in the stellar atmospheres calculations.  We will discuss evolutionary 
times in more detail below.

	Figure \ref{fig:orbparam} shows the cumulative distribution of ELM WD 
masses.  In principle, our surface gravity selection should correspond to uniform 
selection in mass.  In practice, most of the ELM WDs we observe are clumped around 
0.18 \msun.  The discontinuity in the mass distribution is good evidence that, as 
predicted, shell flashes cause $\gtrsim$0.18 \msun\ ELM WDs to evolve much more 
quickly than their lower mass brethren.  If we scale each object by the inverse of 
its evolution time, such that more rapidly evolving ELM WDs contribute a greater 
fraction of the sample, then we obtain a more uniform $M_1$ distribution as seen in 
Figure \ref{fig:orbparam}.

\subsection{Binary Orbital Parameters}

	We adopt the ELM WD binary orbital parameters reported in \citet{brown16a}. 
We include the measured semi-amplitude $k$ and orbital period $P$ in Table 
\ref{tab:clean}, and plot their cumulative distributions in Figure 
\ref{fig:orbparam}.  Orbital periods have a lognormal distribution and a median 
value of 0.21 days.


	ELM WDs dominate the light of the binary systems because the ELM WDs are 
larger in radius and hotter in temperature than their older, more massive WD 
companions.  We thus observe single-lined spectroscopic binaries, and must rely on 
the binary mass function to constrain the mass of the unseen companions.  The binary 
mass function,
	\begin{equation} \frac{P k^3}{2 \pi G} = \frac{(M_2 \sin{i})^3}{(M_1+M_2)^2},
	\end{equation} relates $M_2$ to the measured and derived parameters $P$, 
$k$, and $M_1$ plus the inclination $i$.  We do not know the inclination of 
individual binaries (unless there are eclipses), however we know that the binaries 
were selected by color.  Thus we can assume that the distribution of inclination is 
random in $\sin{i}$ and solve for the underlying distribution of $M_2$. 
\citet{brown16a} show that the best fit to the data comes from a normal distribution 
of $M_2$ with mean 0.76 \msun\ and standard deviation 0.25 \msun\ (see Figure 
\ref{fig:orbparam}).  \citet{andrews14} and \citet{boffin15} find essentially the 
same result using different techniques and an earlier version of the ELM Survey 
data.  Table \ref{tab:clean} includes the estimated $M_2$ for each ELM WD binary 
given any available inclination constraint; we report the median value of $M_2$ 
along with the 0.1587 and 0.8413 percentile values of the $M_2$ distribution.

\begin{figure}          
 \plotone{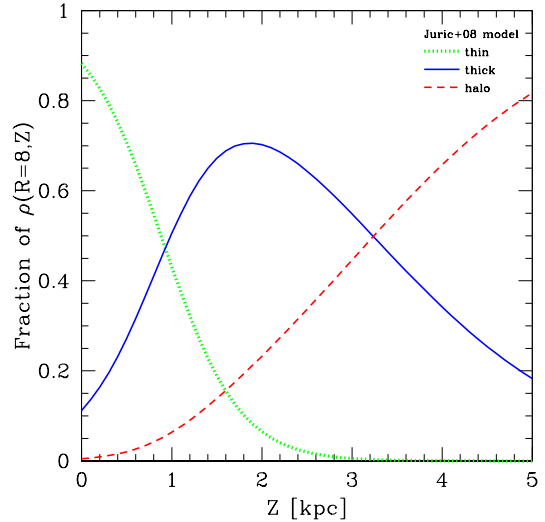}
 \caption{ \label{fig:galmod}
	Galactic stellar density $\rho(R,Z)$ versus height above the disk $Z$ 
calculated at $R=8$ kpc: the thin disk dominates $Z < 1$ kpc (green dotted line), 
the thick disk dominates $Z \sim 2 $ kpc (blue solid line), and the halo dominates 
$Z > 3$ kpc in the \citet{juric08} model. }
 \end{figure}

\subsection{Galactic Stellar Density Models}

	We make use of two Galactic stellar density models in this paper: those of 
\citet{juric08} and \citet{nelemans01b}.  We will use these Galactic stellar density 
models to derive the local space density of ELM WD binaries, and then to infer the 
population of ELM WD binaries over the entire Milky Way.

	\citet{juric08} derive a stellar density model using star counts from SDSS, 
stars that share the same footprint of sky as the ELM WD binaries. \citet{juric08} 
formulate the stellar density model with an exponential distribution for the thin- 
and thick-disk, and a two-axial power-law distribution for the halo:
	\begin{eqnarray} \label{eqn:juric}
\rho (R,Z)_{\rm thin} = & \rho (R_{\odot},0) \exp \left( - \frac{R-R_{\odot}}{2.6} - 
\frac{|Z+Z_{\odot}|}{0.3} \right)  \\
\rho (R,Z)_{\rm thick} = & 0.12 \rho (R_{\odot},0) \exp \left( 
-\frac{R-R_{\odot}}{3.6} - \frac{|Z+Z_{\odot}|}{0.9} \right)  \\
\rho (R,Z)_{\rm halo} = & 0.0051 \rho (R_{\odot},0) \left( \frac{R_{\odot}}{ 
\sqrt{R^2 + (Z/0.64)^2} } \right)^{2.77} ,
	\end{eqnarray} where $R$ and $Z$ are the Galactocentric cylindrical radial 
and vertical distances, respectively, in units of kpc, and $\rho(R,Z)$ is in units 
of kpc$^{-3}$.  We use the \citet{juric08} bias-corrected model parameters, and 
assume that the Sun is located at $(R_{\odot}, Z_{\odot}) = (8, 0.025)$ kpc.

	Figure \ref{fig:galmod} plots the fraction of thin disk, thick disk, and
halo stars as a function of $Z$ in the \citet{juric08} stellar density model,
calculated at the solar position $R_{\odot}=8$ kpc.  This fiducial slice illustrates
that thin disk stars should dominate at $|Z| < 1$ kpc, thick disk stars should
dominate at $|Z| \sim 2 $ kpc, and halo stars should dominate at $|Z| > 3$ kpc.  
The observations sample $R$ and $Z$ in a more complicated way than this simple
slice, however.

\begin{figure*}		
 \plottwo{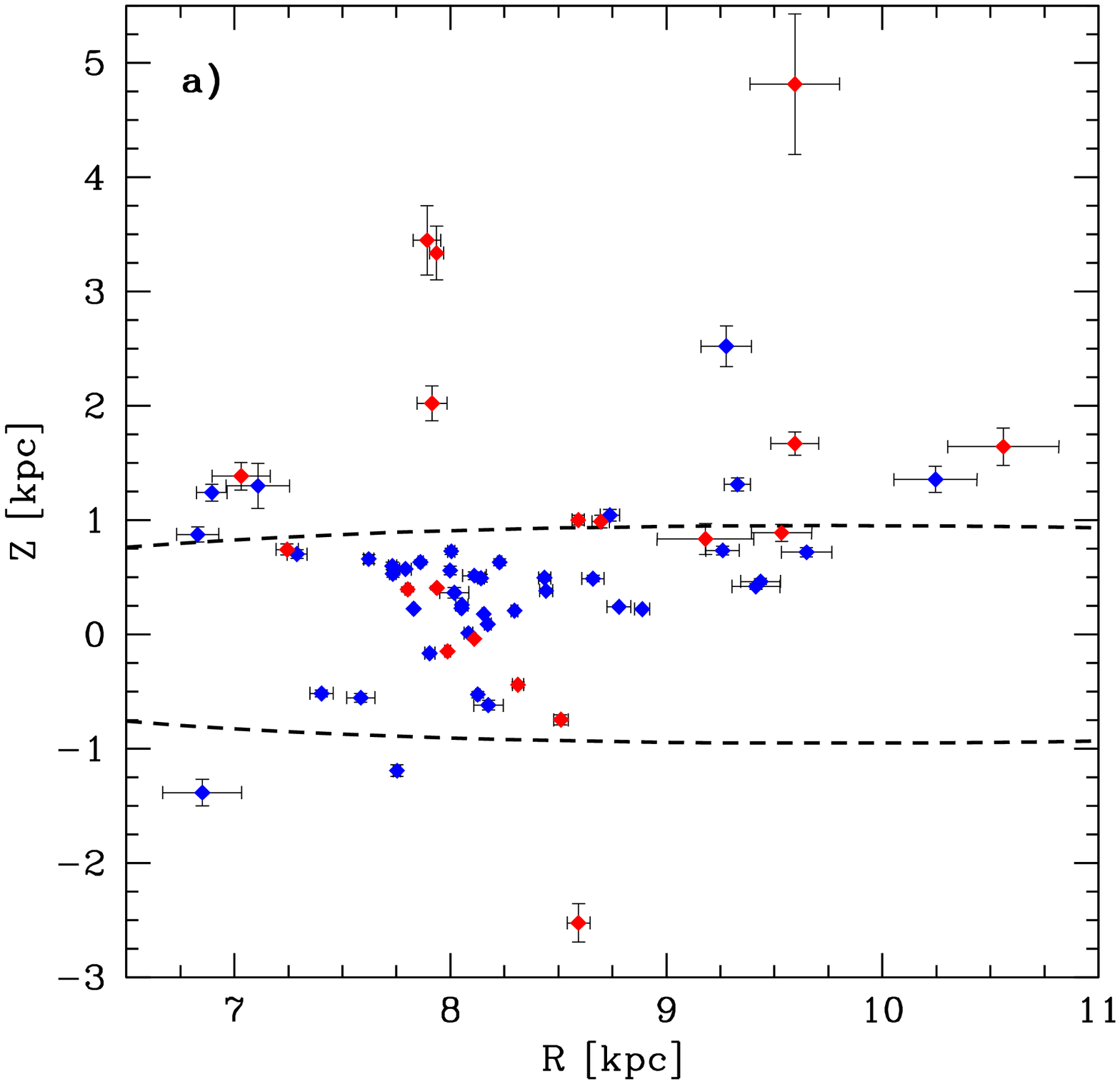}{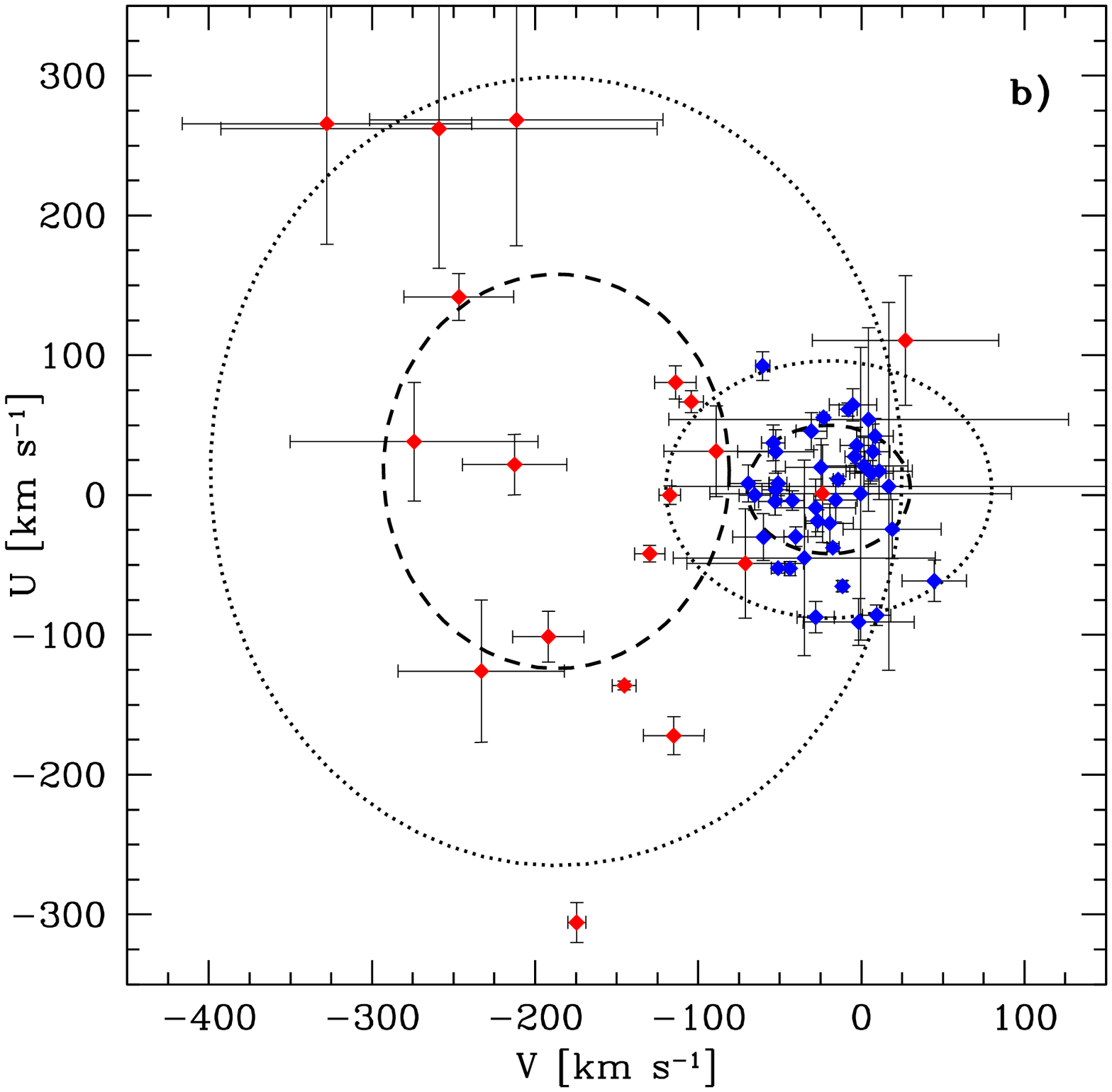}
 \caption{ \label{fig:uvwrz}
	a) Spatial distribution of the clean sample of ELM WD binaries plotted in 
Galactic cylindrical coordinates $R$ and $Z$.  Dashed lines indicate where the ratio 
of disk and halo densities are normalized to 1 for our disk/halo classification 
scheme. Disk objects are marked blue; halo objects red.
	b) Velocity distribution of the clean sample, plotted in Galactic cartesian 
velocity components $U$ (in the direction of the Galactic center) and $V$ (in the 
direction of rotation).  For comparison are the 1$\sigma$ (dashed) and 2$\sigma$ 
(dotted) velocity dispersion thresholds for stellar thick disk and halo populations. 
Symbols same as in a).  }
 \end{figure*}

	Second, we consider the \citet{nelemans01b} disk model commonly used in
AM~CVn studies:
	\begin{eqnarray} 
\rho (R,Z)_{\rm thin} = & 0.98 \rho (R_{\odot},0) \exp \left( - \frac{ R-R_{\odot} 
}{2.5} \right)  {\rm sech} \left(  \frac{ Z+Z_{\odot} }{0.3} \right)^2  \\
\rho (R,Z)_{\rm thick} = & 0.02 \rho (R_{\odot},0) \exp \left( - \frac{ R-R_{\odot}
}{2.5} \right)  {\rm sech} \left(  \frac{ Z+Z_{\odot} }{1.25} \right)^2 .
	\end{eqnarray} The major difference with respect to the \citet{juric08} 
model is the use of sech($Z$)$^2$, which declines more quickly than an exponential, 
and a shorter radial scale length.  The scale length and scale heights used by 
\citet{nelemans01b} originate from earlier star count studies summarized by 
\citet{sackett97}; here, we adopt the thick disk scale height used by 
\citet{roelofs07a}.  Using this disk model allows us to compare with past AM~CVn 
studies and contrast with the \citet{juric08} disk model.

\subsection{Disk / Halo Classification}

	Distinguishing between disk and halo ELM WD binaries is important for our
space density and merger rate analysis.  We will henceforth use the term ``disk'' to
refer to any object in the thin or thick disk, $\rho_{\rm disk} = \rho_{\rm thin} +
\rho_{\rm thick}$.  It is difficult for our observations to distinguish between thin
and thick disk objects, however halo objects stand out by their motion.  Some of the
first ELM WD binary discoveries had large $>$200 \kms\ systemic radial velocities, a
strong indication that they belong to the halo \citep{kilic10, brown11a}.  We thus
base our initial disk/halo classifications on kinematics.

	\citet{gianninas15} combine SDSS+USNO-B proper motions \citep{munn04} with
ELM Survey distance estimates and radial velocities to calculate $(U,V,W)$ space
velocities for the ELM WD binaries.  \citet{gianninas15} then determine disk/halo
membership on the basis of the Mahalanobis distance, or the number of standard
deviations, between an ELM WD space velocity and the velocity ellipsoids of the
thick disk and halo,
	\begin{equation} \label{eqn:mahalanobis}
D_{\rm m} = \sqrt{\frac{(U-\langle U \rangle)^{2}}{\sigma_{U}^{2}} + 
\frac{(V-\langle V \rangle)^{2}}{\sigma_{V}^{2}} + \frac{(W-\langle
W \rangle)^{2}}{\sigma_{W}^{2}}}
	\end{equation} where $(\langle U \rangle, \langle V \rangle, \langle W 
\rangle)$ are the mean velocities and $(\sigma_U, \sigma_V, \sigma_W)$ are the 
velocity dispersions of the thick disk or halo.  We use the velocity ellipsoid 
values from \citet{chiba00}.  A purely kinematic classification suffers some 
ambiguity, however.  As seen in Figure 11 of \citet{gianninas15}, there is an 
overdensity of halo classifications in the disk/halo overlap region around $V=-50$ 
\kms, only seen in that one location of the velocity ellipsoid.  This overdensity is 
more naturally explained as disk objects that happen to be above the 1-$\sigma$ 
thick disk threshold and that are erroneously assigned to the halo.

	We thus introduce an additional consideration -- spatial location -- to the 
selection criteria.  Because we want to prevent extraneous halo classifications in 
the disk-dominated region of the survey, we will require objects closer to the 
disk plane to need more divergent kinematics to be classified as halo objects.  
Specifically, we multiply the ratio of Mahalanobis kinematic distances in Equation 
\ref{eqn:mahalanobis} by the ratio of thick disk to halo space densities.  We 
normalize the ratio of densities to one at the thick disk scale height 0.9 kpc (see 
dashed line in Figure \ref{fig:uvwrz}) to make this a relative weighting that varies 
by about a factor of $e$ over the sample.
	\begin{equation} \label{eqn:diskhalo}
D = \frac{ D_{\rm m}(U,V,W)_{\rm thick} }{ D_{\rm m}(U,V,W)_{\rm halo} } \times 
  \frac{ \rho(R_{\odot},0.9)_{\rm thick} }{ \rho(R,Z)_{\rm thick} }
  \frac{ \rho(R,Z)_{\rm halo} }{ \rho(R_{\odot},0.9)_{\rm halo} },
	\end{equation} where disk objects have $D<1$ and halo objects have $D>1$.  
Adding the spatial consideration changes 10\% of the classifications in the clean 
sample from halo to disk compared to the purely kinematic criteria, and eliminates 
the overdensity of halo classifications clumped around $V=-50$ \kms .

	Because the clean sample contains objects not published in
\citet{gianninas15}, we search for SDSS+USNO-B proper motions for all of the ELM WD
binaries presented here.  Other proper motion catalogs with brighter limiting
magnitudes are not useful because our sample contains faint objects.  Five of the
$g\simeq20$ mag ELM WD binaries have no published proper motion in any catalog.
For these five objects we assume a proper motion of zero and depend solely
on radial velocity; their space velocity can only be larger.  Applying Equation
\ref{eqn:diskhalo} to the clean sample yields 38 (63\%) disk objects and 22 (37\%)
halo objects.  We record the classifications in Table \ref{tab:clean} with disk=1
for disk objects and disk=0 for halo objects.

\subsection{Gravitational Wave Merger Timescale}

	The orbits of ELM WD binaries are shrinking due to gravitational wave 
radiation.  The change in orbital period is observed in the ELM WD binary J0651
\citep{hermes12c}.  The gravitational wave merger timescale of a binary is
	\begin{equation} \label{eqn:gw}
\tau = 47925 \frac{(M_1 + M_2)^{1/3}}{M_1 M_2} P^{8/3} ~{\rm Myr}
        \end{equation} where the masses are in \msun, the period $P$ is in days, and 
the time $\tau$ is in Myr \citep{kraft62}.  Note that the merger time depends much 
more strongly on orbital period than on mass.  The range of ELM WD binary mass is 
also quite limited compared to the range of period.  For the average ELM WD binary, 
$M_1 + M_2 = 1.01 \pm 0.15$ \msun, the merger time at the median $P=0.21$ day is 
$\tau=5$ Gyr.

	For each ELM WD binary, we derive the statistical distribution of $\tau$ 
from its measured $P$ and the derived $M_1$ value and $M_2$ distribution.  We note 
that edge-on orbits with $\sin{i}=1$ provide a minimum $M_2$ and a maximum $\tau$ 
that cannot be exceeded.  We report the median $\tau$ for each binary in Table 
\ref{tab:clean}, along with the 0.1587 and 0.8413 percentile values of the 
distribution.

	The merger times in the clean sample span 1 Myr to over 100 Gyr.  This 
factor of 100,000 in $\tau$ has important implications.  Because the merger rate of 
ELM WD binaries is essentially a harmonic mean of the merger times, the shortest 
merger time systems are the most important for the merger rate calculation.  
Physically, to observe one $\tau=1$ Myr system implies that many more such systems 
must have existed over the $\sim$1 Gyr evolutionary time over which our survey can 
detect an ELM WD.  Conversely, $\tau>13$ Gyr systems exist longer than the age of 
the Universe, thus we should observe all of the ones that are detectable.

	Merger times are strongly linked to orbital period, thus the distribution of 
merger times that we observe also has implications for the initial period 
distribution of the ELM WD binaries.  If ELM WDs continuously detach from the common 
envelope phase at 6 hr periods, for example, half of the clean sample should have 
$P=5$--6 hr and zero should have $P<$1 hr.  The fact that we observe the {\it same} 
number of objects with $P=5$--6 hr as with $P<1$ hr implies that a large fraction of 
ELM WD binaries form at short orbital periods; to observe $P<1$ hr systems requires 
that they are constantly replenished.  We will investigate this point further below.


\subsection{ELM WD Evolutionary Timescales}

	WDs cool and fade with time.  As mentioned before, ELM WDs remain relatively 
luminous over Gyr timescales because of residual hydrogen shell burning in the WD 
atmospheres.  The timescale relevant to this paper, however, is the amount of time 
an ELM WD spends cooling through our color selection region.  We quantify this time 
$t_{obs}$ as follows.

	For each ELM WD evolutionary track, we apply our photometric errors to 
the synthetic colors at each time step in the track and use a Monte Carlo 
calculation to estimate what fraction of simulated observations would fall in our 
color selection region.  The evolutionary tracks contain between 2,000 to 
50,000 time steps and 3 Gyr to 20 Gyr time spans, with an average time step of 1 
Myr in the temperature range sampled by our color selection region. The sum of the 
time steps multiplied by the fractions in our color selection region equals 
$t_{obs}$, the time a WD on a given track could be observed in our survey.

\begin{figure}          
 \plotone{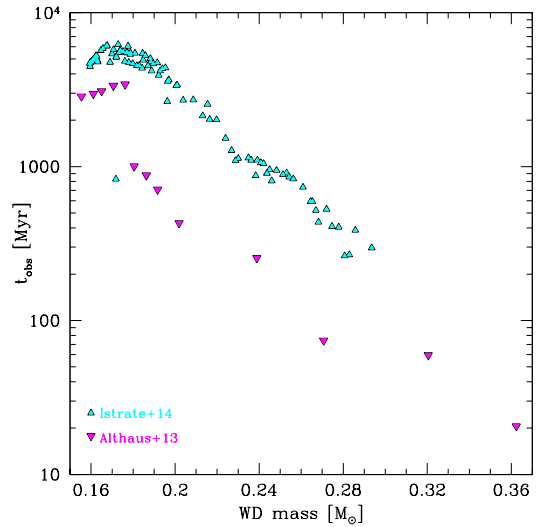}
 \caption{ \label{fig:tobs}
	Timescale $t_{obs}$ that \citet[][cyan]{istrate14} and
\citet[][magenta]{althaus13} ELM WD evolutionary tracks would be observed in our
color selection region.  The break in $t_{obs}$ around 0.18 \msun\ is due to the 
on-set of hydrogen shell flashes in the models. }
 \end{figure}

	The values of $t_{obs}$ are plotted in Figure \ref{fig:tobs} for both
\citet{althaus13} and \citet{istrate14} tracks.  The discontinuity in $t_{obs}$
around 0.18 \msun\ is due to the onset of hydrogen shell flashes in the models.  
Notably, \citet{istrate14} tracks have $t_{obs}$ values that are systematically
$\simeq$5 times longer than the $t_{obs}$ values of the \citet{althaus13} tracks.  
This is likely explained by the absence of gravitational settling in the
\citet{istrate14} models.  We adopt the \citet{althaus13} values in this paper.

	Given the $t_{obs}$ of the models, we estimate $t_{obs}$ for each ELM WD by 
interpolating its measured \teff\ and \logg\ with its errors through the tracks.  
We include the results in Table \ref{tab:clean}.  As before, we report the median 
$t_{obs}$, with the upper and lower uncertainties being the 0.1587 and 0.8413 
percentile values of the distribution.  All $\le 0.18$ \msun\ tracks in the 
\citet{althaus13} models have $t_{obs}=3$ Gyr, and so there is no range to report 
for the lowest mass ELM WDs in the clean sample.

	One implication of the $t_{obs}=3$ Gyr timescale is that many of the ELM WD
binaries we observe represent an integral of the population over the past few Gyr.  
ELM WD binaries with $t_{obs} < \tau$ must remain visible for the full amount of
$t_{obs}$; they will cool out of the survey before they have time to merge.  This
statement applies to ELM WD binaries with approximately $P>0.18$ day periods, about
57\% of the clean sample.

	ELM WD binaries with $t_{obs} > \tau$, on the other hand, represent a 
snapshot of the population.  To observe a $t_{obs} > \tau$ binary requires that it 
form with $P<0.23$ day and evolve over $t_{obs}$ to the period we observe today;  
binaries formed with $P<0.18$ day will merge before they reach $t_{obs}$, which 
means we are lucky to observe them at all.  ELM WD effective temperatures 
corroborate this view:  if the ELM WD binaries merge before they have time to cool 
out of the survey, they must be systematically younger and hotter than the other ELM 
WDs in the survey.  In the clean sample, $P<0.1$ day binaries have median \teff\ = 
16,400 K while $P>0.1$ day binaries have median \teff\ = 11,100 K. The 
Anderson-Darling test gives a p-value of less than 0.0001 that the two \teff\ 
distributions are drawn from the same parent distribution \citep[see 
also][]{gianninas15}.  The significant difference in temperature means short-period 
ELM WD binaries merge before they have time to cool; to see the sample that we 
observe implies that many more short-period ELM WDs must have formed over the past 
few Gyr.

\subsection{Roche Lobe Radius}

	The Roche lobe radius is the effective radius at which an object fills its
gravitational equipotential and begins transfering mass to its companion.  ELM WDs
are always the donor stars in our binaries, because, being degenerate objects, ELM
WDs have larger radii than normal WDs (and also neutron stars) of higher mass.  An
ELM WD will begin mass transfer when its radius exceeds the effective Roche lobe
radius $R_L$,
	\begin{equation} \frac{R_L}{A} = \frac{0.49 q^{2/3}}{ 0.6 q^{2/3} + \ln(1 + 
q^{1/3})}, 
	\end{equation} where $A$ is the orbital separation and $q=M_{\rm ELM}/M_2$ 
is the mass ratio \citep{eggleton83}.  Quantitatively, ELM WDs with $\log{g}=6.5$ 
have 0.04 \rsun\ radii that will exceed $R_L$ at 10 min orbital periods, periods at 
which $\tau\simeq0.6$ Myr.  The gravitational wave merger time is thus broadly 
consistent with the time at which mass transfer will begin for these objects.  This 
is not true for the lowest surface gravity ELM WDs, however.  ELM WDs with 
$\log{g}=5.5$ have $0.13$ \rsun\ radii that exceed $R_L$ at 60 min orbital periods, 
periods at which $\tau\simeq60$ Myr.  The time of merger thus depends on the time 
evolution of ELM WD radii and Roche lobe radii.  We consider both $\tau$ and 
$R_L=R_{\rm ELM}$ as merger conditions in our calculations below.

\section{MERGER RATE ANALYSIS}

\subsection{Space Density of ELM WD Binaries}

	The first step in estimating the merger rate of ELM WD binaries in the Milky
Way is to derive their local space density.  To be clear, we are not deriving the
local space density of all WDs or all ELM WDs, but the density of our clean sample
of ELM WD binaries relevant to the merger rate calculation.  The total number of WD
binaries can only be larger.

	Our approach is to use the modified $1/V_{\rm max}$ method \citep{schmidt75} 
commonly used in WD studies \citep[e.g.][]{rebassa15}.  Stars in the Galaxy are not 
distributed isotropically in a sphere, but are distributed mostly in a disk.  Thus 
the standard approach is to scale the maximum volume $V_{\rm max}$ over which each 
WD can be observed by a Galactic stellar density model,
	\begin{equation}
V_{\rm max} = \int_{l} \int_{b} \int_{r} \frac{ \rho(R,Z) }{ \rho(R_{\odot},0) } 
 dl \cos{b}~db ~r^2 dr,
	\end{equation} where $l$ is Galactic longitude, $b$ is Galactic latitude, 
$r$ is heliocentric distance.  The stellar density model $\rho(R,Z)$ depends on $R$ 
and $Z$ which can be derived from $l$, $b$, and $r$.  As a practical matter, we 
integrate the volume for each WD numerically.  The volume depends on both our survey 
apparent magnitude limits $m_{\rm lim}$ and the SDSS imaging footprint, which is a 
complicated function of $l$ and $b$.  We determine $l$ and $b$ coverage from the 
SDSS stripe definition catalog, clipped by our reddening limit $E(B-V)<0.1$.  We 
observe ELM WDs of different absolute magnitude $M_g$ to different depths $r_{\rm 
lim} = 10^{(m_{\rm lim} - M_g)/5 -2}$ (kpc), where $m_{\rm lim}=15$ and 20.  The 
local space density is then the sum of $1/V_{\rm max}$ over the appropriate sample 
of ELM WD binaries.

	Applying the modified $1/V_{\rm max}$ method to the disk sample, we estimate
a local space density of 160 kpc$^{-3}$ disk ELM WD binaries using the
\citet{juric08} disk model, and 300 kpc$^{-3}$ disk ELM WD binaries using the
\citet{nelemans01b} disk model.  These are observed space densities, corrected only
for survey completeness.  We consider evolutionary time and merger time corrections
in the next subsection.  The factor of two difference between the two estimates is
explained by the shorter thick disk radial scale length and the use of sech($Z$)$^2$
in the \citet{nelemans01b} model, both of which cut off the effective disk volume
and thus boost the local density.  We consider this factor of two the uncertainty in
the space density estimate.  Henceforth, we adopt the \citet{juric08} model-derived
values in our discussion unless making comparison to AM~CVn stars.

	Applying the modified $1/V_{\rm max}$ method to the halo sample, we estimate
a local space density of 6 kpc$^{-3}$ halo ELM WD binaries using the \citet{juric08}
halo model.  This density is 4\% of the disk value, about the same percentage as the
relative normalization of the halo with respect to the thick disk in the
\citet{juric08} models.  The density ratio of disk to halo ELM WD binaries that we
derive from the observations thus appears sensible given the assumed Galactic model.

\begin{figure}          
 \includegraphics[width=3.3in]{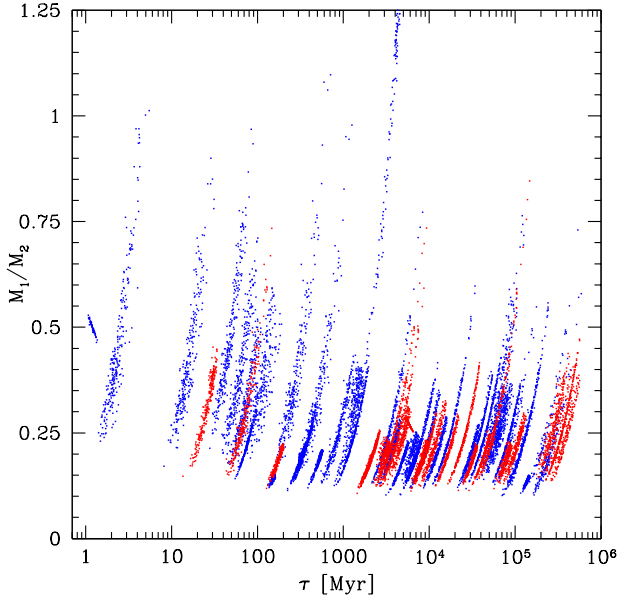}
 \caption{ \label{fig:tau} 
	Gravitational wave merger time $\tau$ versus mass ratio for the clean sample
of ELM WD binaries.  Each dot represents a Monte Carlo calculation accounting for
the observational uncertainties in $M_1$, $P$, and $k$ plus the statistical $M_2$
constraint given any other information (i.e.\ eclipses).  Disk objects are blue,
halo objects are red.}
 \end{figure}

\subsection{Merger Rate of ELM WD Binaries: Reverse Approach}


	We now come to the crux of the problem:  the merger rate of ELM WD binaries
in the Milky Way.  We estimate the merger rate in two ways.  In this subsection, we
derive a merger rate from the observed sample by correcting for the ELM WD binaries
that have cooled or merged over the past Gyr.  This is the approach used in
\citet{brown11a}, however we have a number advantages over our previous work.  We
now have orbital inclination constraints for a number of individual systems:  2
systems are eclipsing \citep{brown11b, kilic14b}, 8 have ellipsoidal variations
\citep{kilic11b, hermes12a, hermes14}, and many more have X-ray and/or radio
observations that rule out neutron star companions \citep{kilic11a, kilic12a,
kilic14b}.  As described above, ELM WD evolutionary models allow us to estimate the
distribution of $t_{obs}$ for each object, and the large sample size allows us to
statistically constrain $M_2$ and thus $\tau$.  Figure \ref{fig:tau} plots the
distribution of $M_1/M_2$ versus $\tau$ relevant to this discussion, colored by disk
and halo classification.

	Given the merger and evolutionary time of each ELM WD binary, we estimate 
the total number of ELM binaries that must have merged or evolved out of the sample 
in the last 1 Gyr.  For example, J0651 has $\tau=1$ Myr.  While it is possible that 
J0651 is a fluke observation, it is not the only short merger time system in our 
sample:  we also observe ELM WD binaries with $\tau=2$, 15, and 27 Myr.  We will 
thus assume that the existence of one $\tau=1$ Myr binary implies that $1/\tau=1000$ 
such objects merged in the last 1 Gyr.  In other words, we assume 
that ELM WD binaries are formed at a constant rate over the past Gyr.  Other 
objects in our sample will cool out of our Survey before they merge.  
Mathematically, each ELM WD binary in our sample corresponds to $N_{\rm corrected}$ 
systems given by the maximum of the $1/\tau$ and $1/t_{obs}$ times.

	Applying these corrections to the disk sample, the local space density of 
disk ELM WD binaries that formed in the last 1 Gyr is 5600 kpc$^{-3}$.  As a sanity 
check, we compare with the local space density implied by the nearest ELM-like WD 
binary.  WD~1242$-$105 is a 0.39 \msun\ + 0.56 \msun\ double degenerate binary with 
a 0.74 Gyr merger time at a distance of 39 pc \citep{debes15}.  Assuming that 66\% 
-- 78\% of all WDs are known within 40 pc \citep{limoges15}, WD~1241$-$105 implies a 
local space density of 5200 kpc$^{-3}$ -- 6100 kpc$^{-3}$ ELM-like WD binaries, 
consistent with our estimate.

	Finally, we integrate the total number of disk ELM WD binaries over the 
Galactic disk model and divide by 1 Gyr to obtain an estimated merger rate of 
$3\times10^{-3}$ yr$^{-1}$.  The halo is less well constrained because the halo 
model is a power law instead of an exponential.  If we integrate the halo model out 
to 100 kpc, the estimated merger rate of halo ELM WD binaries is $3\times10^{-4}$ 
yr$^{-1}$.

	The merger rate of disk ELM WDs is 75 times larger than our previous 
estimate in \citet{brown11a} for two reasons: our sample is 5 times larger than 
before, and we have discovered systems with order-of-magnitude faster merger times.  
The three disk ELM WD binaries with $P<0.028$ day ($1<\tau<15$ Myr) contribute 90\% 
of the corrected number of ELM WD binaries that formed in the last Gyr.  By 
comparison, the 10 systems with $0.042<P<0.08$ day ($45<\tau<360$ Myr) contribute 
only 7\% of the corrected number of disk ELM WD binaries that formed in the last 
Gyr.  Our result is thus sensitive to the shortest orbital period systems in the 
sample.  In the most extreme example, the merger rate drops by a factor of 2.5 if we 
drop the $\tau=1$ Myr system J0651 from the sample.  Yet J0651 exists, as do 
the other $P<0.028$ day systems.

\subsection{Merger Rate of ELM WD Binaries: Forward Approach}

	We estimate the merger rate of ELM WD binaries with a second approach:  
forward-modeling to match the observed distributions.  A forward-modeling approach 
allows us to explore distributions instead of individual objects, and to explore our 
sensitivity to different input assumptions.  The results remain linked to the 
previous estimate, however, because we normalize the model to the observed ELM WD 
binary space density.

\subsubsection{Model}

	The basic idea of our model is to generate ELM WD binaries over a series of
time steps, evolve their orbital periods with gravitational wave radiation, and then
``observe'' the final distribution with the ELM WD evolutionary tracks.  The most
important input is the initial period distribution.  It is difficult to find an
initial period distribution that yields simulated observations that match the actual
observations of $P$ and $k$.  We infer the merger rate by counting the number of
simulated WD ELM binaries that remain observable compared to the number that merged 
over the last 1 Gyr of the simulation.

	In detail, we start by adopting a Galactic star formation rate history.  
Solar neighborhood studies find an approximately constant \citep{cignoni06} or a
modestly rising \citep{tremblay14} star formation rate history for the Galactic
disk.  Whether the star formation rate is increasing, decreasing, or constant over
the last 5 Gyr makes very little difference to the simulations, however, because the
ELM WD binary merger rate is dominated by those systems that form in the last Gyr in
our model.  We adopt a constant formation rate for simplicity.  In our model, we
generate 100 ELM WD binaries at every 1 Myr timestep over the 5 Gyr run of our
simulations.  This can also be thought of as 5,000 bursts of ELM WD binary
formation spread uniformly over time.

	We generate simulated ELM WD binaries by drawing values for $M_1$, $M_2$,
$P$, and $i$ as follows.  We randomly draw $M_1$ from the observed disk or halo
$M_1$ distribution, corrected for $1/t_{obs}$ (green line in Figure
\ref{fig:orbparam}) so that we draw ELM WD masses in the correct proportion.  To
match the observed $M_1$ distribution requires that we draw more short $t_{obs}$ ELM
WDs than long $t_{obs}$ ELM WDs.

	We randomly draw $M_2$ from a normal distribution with mean 0.76 \msun\ and 
dispersion 0.25 \msun, as described above, and impose 0.01 \msun\ and 3 \msun\ 
limits on the allowed range.  We tried exploring different $M_2$ distributions, such 
as the WD mass distribution observed by SDSS \citep{kepler07}, however this has  
only a 10\% effect on merger times and rates because $\tau$ is driven by orbital 
period (Equation \ref{eqn:gw}).  Limiting $M_2$ to $<$1.4 \msun\ also makes little 
difference, because 1.4 \msun\ is 2.6-$\sigma$ from the mean and affects only 0.5\% 
of $M_2$ draws.  After drawing $M_2$, we draw inclination from a random $\sin{i}$ 
distribution.

\begin{figure}          
 \plotone{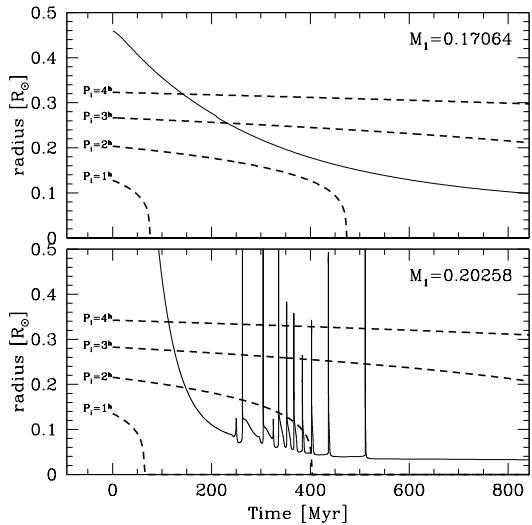}
 \caption{ \label{fig:althaus} 
	Theoretical ELM WD radius (solid line) versus time for 0.17064 \msun\ and 
0.20258 \msun\ evolutionary tracks \citep{althaus13} compared to approximate Roche 
lobe radii $R_L$ (dashed lines) calculated assuming gravitational wave energy loss 
for $M_2=0.76$ \msun\ and four different initial periods $P_i$.  Note the shell 
flashes present in the 0.20258 \msun\ track and absent in the 0.17064 \msun\ track.  
We start our simulated binaries at the moment when $R_{\rm ELM} < R_L$, in 
other words, by shifting the dashed lines to the right in this plot. }
 \end{figure}

	Finally, we randomly draw $P$ from a lognormal distribution,  
	\begin{equation} f_P (P; \mu_P, \sigma_P^2) = \frac{1}{P\sqrt{2\pi}\sigma_P} {\rm 
exp}\left( -\frac{ (\ln{P} - \mu
_P)^2 }{ 2\sigma_P^2 } \right),
	\end{equation} where $P$ is the period, $\mu_P$ is the lognormal mean, and 
$\sigma_P$ is the standard deviation.  The lower bound on $P$ depends on the Roche 
lobe radius, and is thus linked to the evolutionary tracks.  When we adopt $M_1$, 
we adopt a track.  The \citet{althaus13} tracks begin at the moment the ELM WD 
progenitor detaches from the common envelope, which is $P>1$ day (or $\tau>300$ Gyr) 
\citep{althaus13}.  The tracks are thus in conflict with the observed 
ELM WD binaries, which have shorter periods.  Our solution is to find the earliest 
time in a track at which $R_{\rm ELM} < R_L$, and start the calculation from there.  
If $P$ is too small to be allowed at any time in a track, we note the failure and 
re-draw $P$.  This approach presumes that ELM WDs are unchanged by different 
detachment times, but is a self-consistent way of dealing with ELM WD radii.
	Figure \ref{fig:althaus} illustrates the issue.  In solid lines, we plot ELM 
WD radius versus time for two representative tracks from \citet{althaus13}.  In 
dashed lines, we plot the ELM WD Roche lobe radius versus time assuming 
gravitational wave energy loss for $M_2=0.76$ \msun\ and four different initial 
periods $P_i$.  We ignore the brief excursions in $R_{\rm ELM}$ due to shell 
flashes in all of these calculations.

	For each simulated binary, we start our model calculations at the time in a
track when $R_{\rm ELM} < R_L$.  We then evolve the simulated binary forward in time
with gravitational wave energy loss,
	\begin{equation} \frac{dP}{dt} = 7.935\times10^{-6} \frac{M_1 M_2}{ (M_1 +
M_2)^{1/3} } P^{-5/3} ~({\rm days ~ Myr^{-1}}) \end{equation} where $P$ is in days,
$M_1$ and $M_2$ are in \msun, and $dP/dT$ is in days Myr$^{-1}$.  Every 1 Myr time
step we draw additional binaries and update the orbital parameters of the old ones.  
We cease updating a simulated binary when its period becomes short enough that
$R_{\rm ELM}$ from the track exceeds $R_L$, and we count it ``merged.'' We continue
these calculations for 5 Gyr worth of time steps.

\begin{figure*}		
 \plotone{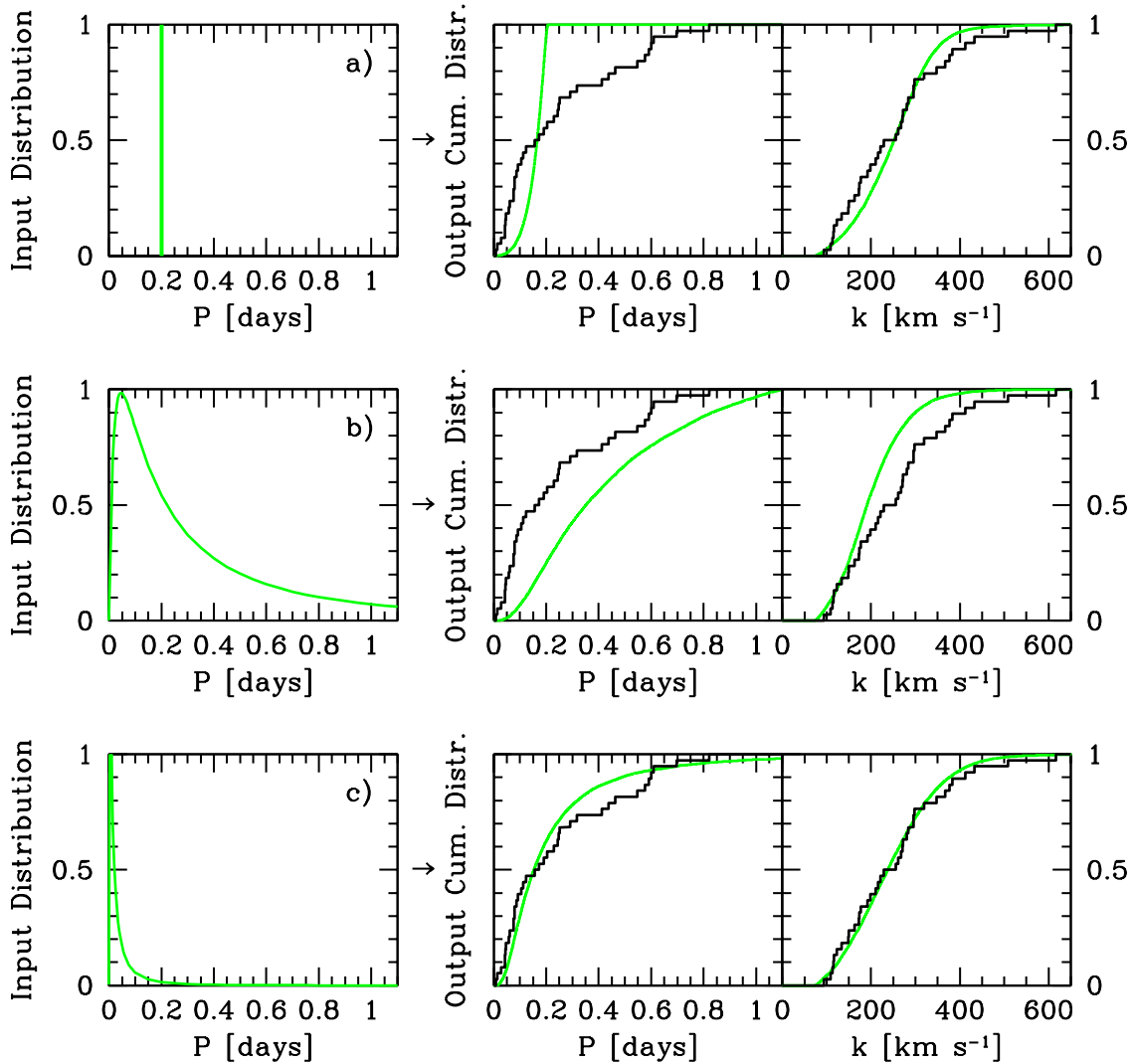}
 \caption{ \label{fig:mct1p} 
	We plot three input period distributions (lefthand panels) and the resulting 
output $P$ and $k$ cumulative distributions evolved by our model (green lines, 
righthand panels) compared to the observations for disk ELM WD binaries (black 
lines).  The bottom row shows the best fit, for input lognormal mean $\mu_P =
-4$ day and standard deviation $\sigma_P = 1.5$. }
 \end{figure*}

	We consider ELM WD \teff\ only at the end of the calculations, when we 
``observe'' the final set of simulated binaries.  For each simulated binary, we look 
up the ELM WD's color given its age along its evolutionary track.  Most of the 
simulated binaries merge or cool out of our color selection before the end of the 
calculations, but a subset remain observable.  We evaluate our choice of input 
parameters by comparing the ``observed'' simulated distributions of $P$ and $k$ to 
the actual observations using the Anderson-Darling test (Figure \ref{fig:mct1p}).

\subsubsection{Results}

	To explain the observed short-period binaries in the disk sample without 
over-producing long-period binaries requires that most ELM WD progenitors detach 
from the common envelope phase with $<1$ hr orbital periods.  Figure \ref{fig:mct1p} 
illustrates the result.  The lefthand column shows three input $P$ distributions.  
The righthand columns show the resulting output $P$ and $k$ distributions evolved by 
our model (in green), compared to the observed disk ELM WD binaries (in black).  
The first input distribution is a test case:  all simulated binaries start with 
$P=0.2$ days.  Such binaries will merge in a few Gyr and are observable at any 
end state, since $t_{obs} \simeq \tau$, however the simulation shows that only 1\% 
would be seen at $<1$ hr periods.  The second (middle) input distribution is a more 
realistic case:  we start with the presently observed lognormal $P$ distribution.  
This input distribution over-produces long-period binaries, however.  Long-period 
binaries have tiny $dP/dt$ and remain observable for the full length of $t_{obs}$.  
We must limit the initial number of long-period binaries to match the observed ratio 
of short- and long-period binaries.

	The best match to the disk ELM WD binaries comes from an input lognormal 
distribution with mean $\mu_P = -4$ day ($e^{-4 ~{\rm days}}$ is about 30 min) 
and standard deviation $\sigma_P = 1.5$.  We identify this match by searching 
over a broad grid $-5<\mu_P<-1$, $0.5<\sigma_P<2.0$, and then refining the grid 
around the best values. As seen in Figure \ref{fig:mct1p}c, the best matching input 
distribution is able to produce an observable number of short-period binaries 
without over-producing long-period binaries; 6\% of the initial draws fail the Roche 
lobe radius test and must be re-drawn.  Importantly, this model predicts that 30 
disk ELM WD binaries merge in the last 1 Gyr for every presently observable system.  
Our observed disk ELM WD sample, integrated over the \citet{juric08} disk model 
($0<R<\infty, -\infty<Z<\infty$), implies there are about $1\times10^5$ observable 
disk ELM WD binaries in the Galaxy.  Multiplying the two numbers together and 
dividing by 1 Gyr yields an estimated disk ELM WD binary merger rate of 
$3\times10^{-3}$ yr$^{-1}$, in agreement with our previous estimate.

	It is unclear whether a constant ELM WD binary formation rate is appropriate 
for the halo, however our model provides an excellent fit to the observed halo 
sample.  Unlike the disk, the observed halo sample contains no rapidly evolving, 
$M_1>0.21$ \msun\ objects.  The best match to the halo observations is for an input 
lognormal distribution with mean $\mu_P = -2$ days (about 3 hrs) and standard 
deviation $\sigma_P = 1.3$, which results in 1.2 halo ELM WD binary mergers in the 
past 1 Gyr for every presently observed system.  Integrating the \citet{juric08} 
halo model to $R=100$ kpc implies there are about $6\times10^5$ observable ELM WD 
binaries throughout the halo.  Multiplying the two numbers together and dividing by 
1 Gyr yields an estimated halo ELM WD binary merger rate of $7\times10^{-4}$ 
yr$^{-1}$, about twice as large as our previous halo estimate.

\subsection{Uncertainty}

	Our merger rate estimates rely on many assumptions, however some matter more 
than others. Changing the disk/halo classification of any ambiguous system alters 
merger rates by $<$1\%, because the ambiguous systems all have long $\tau>200$ Myr 
merger times. The assumed $M_2$ distribution \citep[see][]{brown16a} changes the 
rates by 10\%, because $\tau$ depends only weakly on mass.  The assumed star 
formation history also changes the rates by less than 10\%.

	A major source of uncertainty comes from the small number of short-period
systems that drive the merger rate.  In the reverse approach, three $\tau<15$
Myr systems account for 90\% of the disk merger rate.  If we adopt $\sqrt{N}$ as the
uncertainty of observing the three binaries that dominate the result, the disk
merger rate uncertainty is 60\%.  Bootstrap re-sampling yields the same result,
a 60\% variance in merger rate.  The halo ELM WD binary sample has one $\tau = 25$
Myr system that accounts for half of the halo merger rate; bootstrap re-sampling the
halo sample also yields a 60\% variance in merger rate.  In the forward approach,
bootstrap re-sampling yields a 50\% variance in merger rate.

	Another major source of uncertainty is the observable lifetime of ELM
WDs.  The choice of ELM WD evolutionary model has a 2\% effect on the merger
rate in the reverse approach, because $t_{obs}$ is 20--200 times larger than
$\tau$ for the systems that drive the merger rate calculation.  The forward
approach is more sensitive to $t_{obs}$, however, because the forward approach tries
to match the observed number ratio of short- and long-lived binaries.  If ELM WDs
have larger $t_{obs}$ as in the \citet{istrate14} tracks, then the number of
simulated mergers in the last Gyr is a smaller fraction of the present number of
observable systems.  The \citet{istrate14} tracks lower the merger rate by 60\% in
the forward approach, a difference that we adopt as the effective uncertainty of the
evolutionary tracks.  The larger $t_{obs}$ times are likely explained by the lack of
gravitational settling in the \citet{istrate14} model atmospheres, however, and so 
we rely on \citet{althaus13} tracks for all of our results.

	A final source of uncertainty comes from the Galactic stellar density
model used to infer the total number of ELM WD binaries in the Galaxy.  Integrating
the \citet{juric08} and \citet{nelemans01b} disk models $0<R<\infty$ and
$-\infty<Z<\infty$ yields total numbers that differ by a factor of 1.8.  We
adopt this factor of 1.8 as the uncertainty of the stellar density models.  
We can ignore this uncertainty if we compare with studies that use the
identical Galactic stellar density model.

	Summing the above uncertainties in quadrature, our merger rate uncertainty 
is 80\% (a factor of 1.8) without the Galactic stellar density uncertainty, and 
110\% (a factor of 2.1) with it.  Of course the merger rate is a lower limit to the 
full WD binary merger rate, because we observe only a narrow range of all WD 
binaries.

\section{DISCUSSION}

	Binary population synthesis models have long estimated a total Milky Way
WD+WD merger rate of $(2-5)\times10^{-2}$ yr$^{-1}$, most of which is predicted to
be He+He and He+CO WD mergers \citep{iben90, han98, nelemans01b}.  The ELM WD binary
(He+CO) merger rate we infer from observations is about 10\% of total WD+WD merger
rate, and thus consistent with the model predictions.  We will now link the ELM WD
binary mergers to three possible outcomes on the basis of the rates:  1) long-lived
stable mass-transfer binaries, 2) explosions, and 3) mergers \citep{kilic10}.  
We discuss the disk ELM WD binary merger rate for simplicity; adding the halo
objects increase the merger rate by approximately 10\%, which is less than the
factor of two uncertainty.

\subsection{AM CVn Systems}

	AM~CVn systems are binaries in which a CO WD is accreting helium from a
companion.  One formation channel for AM~CVn systems is a He+CO WD binary, in other
words, an ELM WD binary \citep{solheim10}.  If He+CO WD binaries evolve into stable
mass transfer systems, then population synthesis models predict that the AM~CVn
formation rate is $6.8\times10^{-3}$ yr$^{-1}$ \citep{nelemans01b}.  If He+CO WD
binaries do not evolve into stable mass transfer systems, then other formation
channels dominate and the AM~CVn formation rate is predicted to be
$1.1\times10^{-3}$ yr$^{-1}$ \citep{nelemans01b}.

	Observers have systematically searched for AM~CVn systems in the SDSS and
found 54 times fewer AM~CVn systems than predicted by the optimistic population
synthesis model \citep{roelofs07b, carter13}.  The uncertainty in the space density
of the magnitude-limited AM~CVn sample is 60\% \citep{roelofs07b, carter13}.  The
observations thus imply an AM~CVn formation rate of $(1.3\pm0.8)\times10^{-4}$
yr$^{-1}$.

	The merger rate of disk ELM WD binaries, observed in the same footprint of
sky and using the same \citep{nelemans01b} Galactic stellar density model
parameters, is $5\times10^{-3}$ yr$^{-1}$, 40 times larger than the formation rate
of AM~CVn systems.  This is the ELM WD binary merger rate using the
\citet{nelemans01b} stellar density parameters, the same parameters used in the
AM~CVn observational studies and the theoretical population synthesis models.  The
uncertainty in our ELM WD binary merger rate is a factor of 1.8.  Thus the merger
rate of ELM WD binaries significantly exceeds the formation rate of AM~CVn at over
3-$\sigma$ confidence.  Stated another way, the merger rate of ELM WD binaries is at
least 8 times greater than the formation rate of AM~CVn at the 99\% confidence
level.

\subsection{Underluminous Supernovae}

	If He+CO WD mergers instead result in explosions, they should theoretically
appear as ``underluminous'' supernovae \citep[e.g.][]{bildsten07, guillochon10,
dan11}.  We adopt SN~2008ha as an example of an underluminous supernova.  
\citet{foley09} estimate that SN~2008ha-like events occur at 2\%--10\% of the type
Ia supernovae rate.  From a survey of 1,000 nearby supernovae, \citet{li11}
calculate that the type Ia supernovae rate is $(5.4\pm0.1)\times10^{-3}$ yr$^{-1}$
in Milky Way-like galaxies.  Thus the underluminous supernovae rate in the Milky Way
is approximately $(1-5)\times10^{-4}$ yr$^{-1}$.

	The merger rate of disk ELM WD binaries is 3$\times10^{-3}$ yr$^{-1}$
assuming \citet{juric08} stellar density parameters, or 6--30 times larger than the
underluminous supernovae rate.  The comparison is fraught with uncertainty, however,
given the different manner in which the rates are estimated.  Adopting a factor of
2.1 uncertainty for the ELM WD binary merger rate, the merger rate of ELM WD
binaries exceeds the underluminous supernova rate at 2.9-$\sigma$ confidence.  Thus
it is unlikely that all ELM WD binary mergers explode, but the conclusion is
marginally significant.

\subsection{R Coronae Borealis Stars}

	R~CrB stars are supergiant stars with unusual carbon-rich atmospheres, 76 of
which are known in the Milky Way \citep{tisserand13}.  One formation channel for
R~CrB is the merger of He+CO WD binaries \citep{paczynski71, webbink84}.  Recent
work on R~CrB formation estimates a Milky Way R~CrB formation rate of
$(2-3)\times10^{-3}$ yr$^{-1}$ \citep{zhang14, karakas15}.
	The merger rate of disk ELM WD binaries is statistically identical to the 
R~CrB formation rate.

\subsection{He+CO WD = Unstable Mass Transfer}

	Our ELM WD binary merger rate is only a lower limit to the Milky Way's full
He+CO WD merger rate.  We conclude that ELM WD binaries can be the progenitors of
all observed AM~CVn systems and possibly underluminous supernovae.  The majority of
ELM WD binaries, however, must go through unstable mass transfer and merge, e.g.
into extreme helium stars, R~CrB stars, He-rich subdwarfs, or single massive white
dwarfs.

	The conclusion that ELM WD binaries experience unstable mass transfer and 
merge is surprising given that unstable mass transfer is only assured for mass 
ratios $M_1 / M_2 > 2/3$.  Below that threshold there is a large region of parameter 
space in which the stability of mass transfer is ambiguous, and depends primarily 
on the strength of spin-orbit coupling \citep{marsh04}.  \citet{kremer15} recently 
publish detailed angular momentum calculations of merging WD binaries that assumes
strong spin-orbit coupling, and essentially all of their theoretical ELM WD binaries 
evolve into AM~CVn systems.  The observed rates do not match up, however.

	Our observations imply that spin-orbit coupling in He+CO WD binaries is very
weak.  If the accretor (which spins-up due to the incoming matter stream) is unable
to transfer angular momentum back to the orbit of the ELM WD on a fast enough
timescale, the binary orbit shrinks, mass transfer grows and becomes unstable, and
the WDs will merge \citep{marsh04}.  An alternative explanation may lie in the
initial phase of hydrogen mass transfer.  \citet{shen15} argues that the hydrogen
mass transfer will produce a nova-like outburst that will shrink the binary orbit
and drive the mass transfer rate unstable, causing a merger.  Either way, our
observational constraints on the ELM WD merger rate provide, for the first time,
compelling evidence that mass transfer in most He+CO WD binaries is unstable.

\section{CONCLUSION}

	The goal of this paper is to study the merger rate of ELM WD binaries in the 
Galaxy.  We start by drawing a clean sample of ELM WD binaries from the ELM Survey, 
a spectroscopic survey of color-selected low mass WD candidates.  We classify 63\% 
of the clean sample as disk and 37\% as halo on the basis of kinematics and spatial 
position.  We derive the local space density of ELM WD binaries using the standard 
$1/V_{\rm max}$ method and a Galactic stellar density model, and then use this model 
to estimate that $10^5$ ELM WD binaries presently fall in our color-selection in the 
Galaxy.

	There are two important timescales to consider.  First is the evolutionary
time, $t_{obs}$, that an ELM WD spends cooling through our color-selection.  ELM WDs
with $<0.18$ \msun\ are predicted to have long 3 Gyr evolutionary times, while ELM
WDs with $>0.18$ \msun\ experience shell flashes and cool much more quickly
\citep{althaus13, istrate14}.  The observed distribution of ELM WD masses, clumped
around 0.18 \msun\ with a tail to higher mass, corroborates the predicted
evolutionary timescales.

	The second important timescale is the gravitational wave merger time,
$\tau$, that is a strong function of orbital period.  Observed ELM WD binaries have
gravitational wave merger times that range from 1 Myr to over 100 Gyr.  The shortest
merger time systems drive the merger rate calculation, and this raises a concern: a
recent paper about the gravitational wave properties of the $\tau=1$ Myr system
J0651 wryly points out that ``the chances of observing an eclipsing high-$f$
[short-period] binary within 1 kpc is almost zero'' \citep{shah14}.  Yet J0651 and
the other short-period ELM WD binaries exist.  We conclude that the only way to
match the observed number of short-period systems is if most of the progenitor He+CO
WD binaries detach from the common envelope phase at $P<1$ hr periods.

	We estimate the ELM WD binary merger rate in two ways.  First, we correct
the observed sample for the number of objects that must have merged or evolved out
of the sample in the last 1 Gyr.  This approach yields an ELM WD binary merger rate
of $3\times10^{-3}$ yr$^{-1}$ for the disk and a 10 times lower rate for the halo,
with an uncertainty of a factor of 2.  Second, we forward model to match the
observed distributions.  The results are insensitive to the choice of Galactic star
formation rate history and secondary mass distribution, but are sensitive to the
input orbital period distribution and choice of WD evolutionary tracks.  The
model that best describes the observations yields the same answer as before,
$3\times10^{-3}$ yr$^{-1}$, for disk ELM WD binaries.

	The estimated merger rate of disk ELM WD binaries is 40 times larger than 
the formation rate of AM~CVn systems, marginally larger than the rate of 
underluminous supernovae, and statistically identical to the formation rate of R~CrB 
stars.  On this basis, we conclude that ELM WD binaries can be the progenitors of 
all observed AM~CVn and possibly underluminous supernovae, however the majority of 
ELM WD binaries must merge into single massive objects.

	This conclusion is unexpected given the relatively extreme mass ratios of
the ELM WD binaries.  And since double WD binaries with more massive WDs have mass
ratios closer to one, we conclude that the majority of all double WD binaries must
merge.  In the future, we will explore a more detailed comparison of our
observations with theoretical binary population synthesis models.  The combination
of merger rate, orbital period, and binary mass ratio is potentially a very powerful
constraint on ELM WD binary formation channels and outcomes.  Furthermore, taking
the ELM Survey to its conclusion should double the sample size of ELM WD binaries,
improving our constraints on the important short-period end of the distribution.  
There are many more interesting ELM WD binaries to be found.


\acknowledgments
	This research makes use of the SAO/NASA Astrophysics Data System 
Bibliographic Service.  We thank the referee for a timely and constructive report.  
This work was supported in part by the Smithsonian Institution.  MK and AG 
gratefully acknowledge the support of the NSF and NASA under grants AST-1312678 and 
NNX14AF65G, respectively.

	\clearpage


	\clearpage

\begin{deluxetable}{cccccccc}
\tabletypesize{\scriptsize}
\tablecolumns{8}
\tablewidth{0pt}
\tablecaption{Clean Sample of ELM WD Binaries\label{tab:clean}}
\tablehead{
\colhead{Object}&
\colhead{$P$}&
\colhead{$k$}&
\colhead{$M_1$}&
\colhead{$M_2$}&
\colhead{$t_{\rm obs}$}&
\colhead{$\tau$}&
\colhead{disk}\\
  & (days) & (km s$^{-1}$) & (\msun) & (\msun) & (Gyr) & (Gyr) & 
}
	\startdata
0651$+$2844 & $0.00886 \pm 0.00001$ & $616.9 \pm 5.0$ & $0.247 \pm 0.015$ & $0.49 _{-0.02} ^{+0.02}$ & $0.21 _{-0.05} ^{+0.05}$ & $  0.001 _{-0.0001}^{+0.0001}$ & 1 \\
0935$+$4411 & $0.01375 \pm 0.00051$ & $198.5 \pm 3.2$ & $0.312 \pm 0.019$ & $0.75 _{-0.23} ^{+0.24}$ & $0.05 _{-0.01} ^{+0.02}$ & $  0.002 _{-0.0004}^{+0.0008}$ & 1 \\
0106$-$1000 & $0.02715 \pm 0.00002$ & $395.2 \pm 3.6$ & $0.188 \pm 0.011$ & $0.57 _{-0.07} ^{+0.22}$ & $0.71 _{-0.04} ^{+0.06}$ & $  0.027 _{-0.006} ^{+0.003}$ & 0 \\
1630$+$4233 & $0.02766 \pm 0.00004$ & $295.9 \pm 4.9$ & $0.298 \pm 0.019$ & $0.76 _{-0.22} ^{+0.24}$ & $0.07 _{-0.02} ^{+0.03}$ & $  0.015 _{-0.003} ^{+0.005}$ & 1 \\
1053$+$5200 & $0.04256 \pm 0.00002$ & $264.0 \pm 2.0$ & $0.204 \pm 0.012$ & $0.75 _{-0.23} ^{+0.24}$ & $0.51 _{-0.06} ^{+0.06}$ & $  0.068 _{-0.012} ^{+0.021}$ & 0 \\
0056$-$0611 & $0.04338 \pm 0.00002$ & $376.9 \pm 2.4$ & $0.180 \pm 0.010$ & $0.82 _{-0.11} ^{+0.14}$ & $3.00 _{-2.17} ^{+0.00}$ & $  0.076 _{-0.008} ^{+0.008}$ & 1 \\
1056$+$6536 & $0.04351 \pm 0.00103$ & $267.5 \pm 7.4$ & $0.334 \pm 0.016$ & $0.76 _{-0.22} ^{+0.24}$ & $0.03 _{-0.01} ^{+0.01}$ & $  0.045 _{-0.009} ^{+0.014}$ & 1 \\
0923$+$3028 & $0.04495 \pm 0.00049$ & $296.0 \pm 3.0$ & $0.275 \pm 0.015$ & $0.76 _{-0.21} ^{+0.23}$ & $0.12 _{-0.03} ^{+0.03}$ & $  0.059 _{-0.011} ^{+0.017}$ & 1 \\
1436$+$5010 & $0.04580 \pm 0.00010$ & $347.4 \pm 8.9$ & $0.234 \pm 0.013$ & $0.78 _{-0.19} ^{+0.23}$ & $0.27 _{-0.04} ^{+0.07}$ & $  0.071 _{-0.012} ^{+0.017}$ & 1 \\
0825$+$1152 & $0.05819 \pm 0.00001$ & $319.4 \pm 2.7$ & $0.278 \pm 0.021$ & $0.80 _{-0.19} ^{+0.22}$ & $0.10 _{-0.03} ^{+0.07}$ & $  0.113 _{-0.020} ^{+0.027}$ & 1 \\
1741$+$6526 & $0.06111 \pm 0.00001$ & $508.0 \pm 4.0$ & $0.170 \pm 0.010$ & $1.17 _{-0.03} ^{+0.07}$ &  3                       & $  0.154 _{-0.006} ^{+0.004}$ & 1 \\
0755$+$4906 & $0.06302 \pm 0.00213$ & $438.0 \pm 5.0$ & $0.184 \pm 0.010$ & $0.96 _{-0.10} ^{+0.16}$ & $0.78 _{-0.05} ^{+0.03}$ & $  0.178 _{-0.019} ^{+0.015}$ & 0 \\
2338$-$2052 & $0.07644 \pm 0.00712$ & $133.4 \pm 7.5$ & $0.258 \pm 0.015$ & $0.75 _{-0.23} ^{+0.24}$ & $0.16 _{-0.04} ^{+0.05}$ & $  0.262 _{-0.050} ^{+0.087}$ & 1 \\
2309$+$2603 & $0.07653 \pm 0.00001$ & $412.4 \pm 2.7$ & $0.176 \pm 0.010$ & $0.96 _{-0.10} ^{+0.16}$ &  3                       & $  0.313 _{-0.033} ^{+0.025}$ & 1 \\
0849$+$0445 & $0.07870 \pm 0.00010$ & $366.9 \pm 4.7$ & $0.179 \pm 0.010$ & $0.86 _{-0.14} ^{+0.19}$ &  3                       & $  0.360 _{-0.049} ^{+0.049}$ & 1 \\
0751$-$0141 & $0.08001 \pm 0.00279$ & $432.6 \pm 2.3$ & $0.194 \pm 0.010$ & $0.98 _{-0.01} ^{+0.01}$ & $0.63 _{-0.02} ^{+0.02}$ & $  0.317 _{-0.004} ^{+0.003}$ & 1 \\
2119$-$0018 & $0.08677 \pm 0.00004$ & $383.0 \pm 4.0$ & $0.159 \pm 0.010$ & $0.84 _{-0.05} ^{+0.14}$ &  3                       & $  0.529 _{-0.055} ^{+0.025}$ & 1 \\
1234$-$0228 & $0.09143 \pm 0.00400$ & $ 94.0 \pm 2.3$ & $0.227 \pm 0.014$ & $0.75 _{-0.23} ^{+0.24}$ & $0.31 _{-0.06} ^{+0.08}$ & $  0.476 _{-0.089} ^{+0.157}$ & 1 \\
1054$-$2121 & $0.10439 \pm 0.00655$ & $261.1 \pm 7.1$ & $0.178 \pm 0.011$ & $0.77 _{-0.20} ^{+0.24}$ & $3.00 _{-2.17} ^{+0.00}$ & $  0.832 _{-0.148} ^{+0.212}$ & 1 \\
0745$+$1949 & $0.11240 \pm 0.00833$ & $108.7 \pm 2.9$ & $0.164 \pm 0.010$ & $0.15 _{-0.03} ^{+0.34}$ &  3                       & $  3.91  _{-2.39}  ^{+0.85} $ & 1 \\
1108$+$1512 & $0.12310 \pm 0.00867$ & $256.2 \pm 3.7$ & $0.179 \pm 0.010$ & $0.78 _{-0.20} ^{+0.22}$ &  3                       & $  1.27  _{-0.21}  ^{+0.30} $ & 1 \\
0112$+$1835 & $0.14698 \pm 0.00003$ & $295.3 \pm 2.0$ & $0.160 \pm 0.010$ & $0.74 _{-0.05} ^{+0.15}$ &  3                       & $  2.35  _{-0.30}  ^{+0.13} $ & 0 \\
1233$+$1602 & $0.15090 \pm 0.00009$ & $336.0 \pm 4.0$ & $0.169 \pm 0.010$ & $0.98 _{-0.09} ^{+0.16}$ &  3                       & $  1.95  _{-0.20}  ^{+0.15} $ & 0 \\
1130$+$3855 & $0.15652 \pm 0.00001$ & $284.0 \pm 4.9$ & $0.288 \pm 0.018$ & $0.90 _{-0.12} ^{+0.18}$ & $0.09 _{-0.03} ^{+0.04}$ & $  1.40  _{-0.19}  ^{+0.18} $ & 1 \\
1112$+$1117 & $0.17248 \pm 0.00001$ & $116.2 \pm 2.8$ & $0.176 \pm 0.010$ & $0.75 _{-0.23} ^{+0.24}$ &  3                       & $  3.26  _{-0.60}  ^{+1.03} $ & 1 \\
1005$+$3550 & $0.17652 \pm 0.00011$ & $143.0 \pm 2.3$ & $0.168 \pm 0.010$ & $0.75 _{-0.23} ^{+0.24}$ &  3                       & $  3.62  _{-0.66}  ^{+1.14} $ & 0 \\
0818$+$3536 & $0.18315 \pm 0.02110$ & $170.0 \pm 5.0$ & $0.165 \pm 0.010$ & $0.75 _{-0.23} ^{+0.24}$ &  3                       & $  4.06  _{-0.73}  ^{+1.24} $ & 0 \\
1443$+$1509 & $0.19053 \pm 0.02402$ & $306.7 \pm 3.0$ & $0.201 \pm 0.013$ & $0.99 _{-0.09} ^{+0.15}$ & $0.56 _{-0.11} ^{+0.09}$ & $  3.06  _{-0.32}  ^{+0.27} $ & 0 \\
1840$+$6423 & $0.19130 \pm 0.00005$ & $272.0 \pm 2.0$ & $0.182 \pm 0.011$ & $0.86 _{-0.14} ^{+0.19}$ & $0.84 _{-0.09} ^{+2.16}$ & $  3.76  _{-0.51}  ^{+0.53} $ & 1 \\
2103$-$0027 & $0.20308 \pm 0.00023$ & $281.0 \pm 3.2$ & $0.161 \pm 0.010$ & $0.88 _{-0.12} ^{+0.19}$ &  3                       & $  4.87  _{-0.62}  ^{+0.56} $ & 1 \\
1238$+$1946 & $0.22275 \pm 0.00009$ & $258.6 \pm 2.5$ & $0.210 \pm 0.011$ & $0.87 _{-0.13} ^{+0.19}$ & $0.45 _{-0.05} ^{+0.06}$ & $  4.91  _{-0.67}  ^{+0.62} $ & 0 \\
1249$+$2626 & $0.22906 \pm 0.00112$ & $191.6 \pm 3.9$ & $0.160 \pm 0.010$ & $0.76 _{-0.22} ^{+0.23}$ &  3                       & $  7.51  _{-1.31}  ^{+2.10} $ & 1 \\
0822$+$2753 & $0.24400 \pm 0.00020$ & $271.1 \pm 9.0$ & $0.191 \pm 0.012$ & $0.93 _{-0.11} ^{+0.17}$ & $0.66 _{-0.08} ^{+0.18}$ & $  6.52  _{-0.78}  ^{+0.64} $ & 1 \\
1631$+$0605 & $0.24776 \pm 0.00411$ & $215.4 \pm 3.4$ & $0.162 \pm 0.010$ & $0.79 _{-0.19} ^{+0.23}$ &  3                       & $  8.95  _{-1.51}  ^{+2.01} $ & 1 \\
1526$+$0543 & $0.25039 \pm 0.00001$ & $231.9 \pm 2.3$ & $0.161 \pm 0.010$ & $0.81 _{-0.17} ^{+0.21}$ &  3                       & $  9.07  _{-1.40}  ^{+1.74} $ & 0 \\
2132$+$0754 & $0.25056 \pm 0.00002$ & $297.3 \pm 3.0$ & $0.187 \pm 0.010$ & $1.07 _{-0.07} ^{+0.13}$ & $0.71 _{-0.03} ^{+0.06}$ & $  6.43  _{-0.53}  ^{+0.34} $ & 1 \\
1141$+$3850 & $0.25958 \pm 0.00005$ & $265.8 \pm 3.5$ & $0.177 \pm 0.010$ & $0.92 _{-0.11} ^{+0.17}$ &  3                       & $  8.32  _{-0.97}  ^{+0.79} $ & 0 \\
1630$+$2712 & $0.27646 \pm 0.00002$ & $218.0 \pm 5.0$ & $0.170 \pm 0.010$ & $0.80 _{-0.17} ^{+0.22}$ &  3                       & $ 11.3   _{-1.8}   ^{+2.2}  $ & 0 \\
1449$+$1717 & $0.29075 \pm 0.00001$ & $228.5 \pm 3.2$ & $0.171 \pm 0.010$ & $0.83 _{-0.15} ^{+0.21}$ &  3                       & $ 12.5   _{-1.9}   ^{+2.0}  $ & 1 \\
0917$+$4638 & $0.31642 \pm 0.00002$ & $148.8 \pm 2.0$ & $0.173 \pm 0.010$ & $0.75 _{-0.23} ^{+0.23}$ &  3                       & $ 16.6   _{-2.9}   ^{+5.1}  $ & 1 \\
0152$+$0749 & $0.32288 \pm 0.00014$ & $217.0 \pm 2.0$ & $0.169 \pm 0.010$ & $0.82 _{-0.16} ^{+0.21}$ &  3                       & $ 16.9   _{-2.5}   ^{+2.9}  $ & 0 \\
1422$+$4352 & $0.37930 \pm 0.01123$ & $176.0 \pm 6.0$ & $0.181 \pm 0.010$ & $0.78 _{-0.20} ^{+0.23}$ & $0.81 _{-0.01} ^{+2.19}$ & $ 25.2   _{-4.2}   ^{+6.0}  $ & 0 \\
1617$+$1310 & $0.41124 \pm 0.00086$ & $210.1 \pm 2.8$ & $0.172 \pm 0.010$ & $0.85 _{-0.14} ^{+0.20}$ &  3                       & $ 30.9   _{-4.4}   ^{+4.4}  $ & 1 \\
1538$+$0252 & $0.41915 \pm 0.00295$ & $227.6 \pm 4.9$ & $0.168 \pm 0.010$ & $0.92 _{-0.11} ^{+0.17}$ &  3                       & $ 31.5   _{-3.7}   ^{+3.0}  $ & 0 \\
1439$+$1002 & $0.43741 \pm 0.00169$ & $174.0 \pm 2.0$ & $0.181 \pm 0.010$ & $0.78 _{-0.19} ^{+0.23}$ & $0.84 _{-0.05} ^{+2.16}$ & $ 36.8   _{-6.2}   ^{+8.5}  $ & 1 \\
0837$+$6648 & $0.46329 \pm 0.00005$ & $150.3 \pm 3.0$ & $0.181 \pm 0.010$ & $0.76 _{-0.22} ^{+0.24}$ & $0.83 _{-0.02} ^{+2.17}$ & $ 43.7   _{-7.7}  ^{+12.1} $ & 1 \\
0940$+$6304 & $0.48438 \pm 0.00001$ & $210.4 \pm 3.2$ & $0.180 \pm 0.010$ & $0.90 _{-0.12} ^{+0.18}$ & $0.83 _{-0.01} ^{+2.17}$ & $ 43.9   _{-5.5}   ^{+4.8}  $ & 0 \\
0840$+$1527 & $0.52155 \pm 0.00474$ & $ 84.8 \pm 3.1$ & $0.192 \pm 0.010$ & $0.75 _{-0.23} ^{+0.24}$ & $0.66 _{-0.03} ^{+0.03}$ & $ 57.5   _{-10.5}  ^{+18.4} $ & 0 \\
0802$-$0955 & $0.54687 \pm 0.00455$ & $176.5 \pm 4.5$ & $0.197 \pm 0.012$ & $0.82 _{-0.16} ^{+0.21}$ & $0.58 _{-0.09} ^{+0.10}$ & $ 59.7   _{-9.3}  ^{+10.7} $ & 1 \\
1518$+$1354 & $0.57655 \pm 0.00734$ & $112.7 \pm 4.6$ & $0.147 \pm 0.018$ & $0.75 _{-0.23} ^{+0.24}$ &  3                       & $ 97.8   _{-19.7}  ^{+31.3} $ & 1 \\
2151$+$1614 & $0.59152 \pm 0.00008$ & $163.3 \pm 3.1$ & $0.181 \pm 0.010$ & $0.80 _{-0.19} ^{+0.22}$ & $0.82 _{-0.02} ^{+0.02}$ & $ 81.2   _{-13.2}  ^{+17.9} $ & 1 \\
1512$+$2615 & $0.59999 \pm 0.02348$ & $115.0 \pm 4.0$ & $0.250 \pm 0.014$ & $0.76 _{-0.23} ^{+0.24}$ & $0.19 _{-0.03} ^{+0.04}$ & $ 65.0   _{-12.1}  ^{+20.3} $ & 1 \\
1518$+$0658 & $0.60935 \pm 0.00004$ & $172.0 \pm 2.0$ & $0.224 \pm 0.013$ & $0.83 _{-0.15} ^{+0.20}$ & $0.34 _{-0.05} ^{+0.06}$ & $ 69.8   _{-10.6}  ^{+11.5} $ & 1 \\
0756$+$6704 & $0.61781 \pm 0.00002$ & $204.2 \pm 1.6$ & $0.182 \pm 0.011$ & $0.95 _{-0.10} ^{+0.16}$ & $0.83 _{-0.10} ^{+2.17}$ & $ 79.9   _{-8.7}   ^{+6.9}  $ & 0 \\
1151$+$5858 & $0.66902 \pm 0.00070$ & $175.7 \pm 5.9$ & $0.186 \pm 0.011$ & $0.85 _{-0.15} ^{+0.19}$ & $0.74 _{-0.06} ^{+0.06}$ & $105     _{-15}    ^{+16}   $ & 0 \\
0730$+$1703 & $0.69770 \pm 0.05427$ & $122.8 \pm 4.3$ & $0.182 \pm 0.010$ & $0.76 _{-0.22} ^{+0.24}$ & $0.81 _{-0.01} ^{+0.02}$ & $130     _{-23}    ^{+37}   $ & 1 \\
0811$+$0225 & $0.82194 \pm 0.00049$ & $220.7 \pm 2.5$ & $0.179 \pm 0.010$ & $1.28 _{-0.05} ^{+0.10}$ & $3.00 _{-2.17} ^{+0.00}$ & $141     _{-7}    ^{+4}   $ & 1 \\
1241$+$0633 & $0.95912 \pm 0.00028$ & $138.2 \pm 4.8$ & $0.199 \pm 0.012$ & $0.80 _{-0.18} ^{+0.22}$ & $0.57 _{-0.07} ^{+0.06}$ & $269     _{-45}    ^{+58}   $ & 0 \\
2236$+$2232 & $1.01016 \pm 0.00005$ & $119.9 \pm 2.0$ & $0.186 \pm 0.010$ & $0.77 _{-0.21} ^{+0.23}$ & $0.75 _{-0.01} ^{+0.02}$ & $339     _{-59}    ^{+89}   $ & 0 \\
0815$+$2309 & $1.07357 \pm 0.00018$ & $131.7 \pm 2.6$ & $0.199 \pm 0.021$ & $0.80 _{-0.19} ^{+0.22}$ & $0.67 _{-0.35} ^{+0.11}$ & $366     _{-66}    ^{+84}   $ & 0 
	\enddata
	\tablecomments{$M_1$ and $t_{\rm obs}$ derived from \citet{althaus13} 
models.}
\end{deluxetable}

\end{document}